\documentstyle[12pt]{article}
\begin{document}


\def\beq{\begin{equation}} 
\def\eeq{\end{equation}} 
\def\bea{\begin{eqnarray}} 
\def\eea{\end{eqnarray}} 
\def\bq{\begin{quote}} 
\def\eq{\end{quote}}
\def\ra{\rightarrow} 
\def\lra{\leftrightarrow} 

\parskip 0.3cm 
\pagestyle{empty} 
\begin{flushright} 
hep-ph/yymmxxx \\
{IOA.335/95}\\ 
\end{flushright} 
\vspace*{3cm}
\begin{center} 
{\bf A String Model with $SU(4)\times O(4)\times
 [Sp(4)]_{Hidden}$ Gauge Symmetry }\\ 

\vspace*{2cm} {\bf G. K. Leontaris} \\
\vspace*{0.2cm} 
{\it Theoretical Physics Division} \\
{\it Ioannina University} \\
{\it GR-45110 Greece} \\ 
\vspace*{0.5cm} 
{\bf ABSTRACT} \\ 
\end{center} 
\noindent 
In the four dimensional free  fermionic formulation of the  heterotic 
string, a semi-realistic $SU(4)\times SU(2)_L \times SU(2)_R$ model is  
proposed  with  three  fermion  generations in $(4,2,1)+(\bar 4,1,2)$
representations.  The gauge symmetry of the model breaks to the 
standard gauge group using a higgs pair in the $(4,1,2)+(\bar 4,1,2)$
representations. The massless spectrum includes  exotic fractionally 
charged states  with non -- trivial transformation properties under 
part ($Sp(4)$) of the non -- abelian `hidden' symmetry. Finally
there is a mirror pair in $(4,2,1) , (\bar 4,2,1)$  allowing the 
possibility for an  identical running of $g_{4,L,R}$ couplings
between the string and $SU(4)$ breaking scales. This is of crucial 
importance for a successful prediction of the weak mixing angle.
Potential shortcomings and problems of the construction are analysed 
and possible solutions are discussed.
\vspace*{2cm}
\noindent
\begin{flushleft} 
IOA-335/95 \\
September 95
\end{flushleft} 
\vfill\eject
\pagestyle{plain} 
\setcounter{page}{1}

   One of the most challenging and interesting issues in strings
\cite{.}, is to construct realistic models [2-10], consistent with 
the  low energy theory.  Most of the attempts in this direction 
\cite{mb,aehn,alr,af},  have been concentrated in constructions 
of string  models based on level-one ($k=1$) Kac-Moody algebras. 
At the $k=1$ level,  several obstacles have appeared:
First, unified models based one these constructions do not
contain higgs fields in the adjoint or higher representations, 
therefore, traditional Grand Unified Theories (GUTs), like SU(5) 
and SO(10) could not break down to the Standard Model. 
Attempts to overcome this difficulty, led to constructions
which needed only small higgs representations to break the
symmetry \cite{aehn,alr}.
\footnote{more recent attempts\cite{bf,fin} to overcome
this difficulty have led to $SO(10)\times SO(10)$ or $SU(5)\times
SU(5)$ product groups, where the $SO(10)$ or $SU(5)$ are realized
directly at level 1.}
A second difficulty\cite{fc} that was encountered within the  $k=1$
Kac -- Moody models,  was the appearance of fractionally charged
states other than the ordinary Quarks, in the particle spectrum. 
Such states, --unless they become massive at the string scale--,
they usually create problems in the low energy effective theory.
Indeed, the lightest fractionally charged particle is expected to be
stable. In particular, if its mass lies in the TeV region, then
the estimation of its relic abundances\cite{ra} contradicts the upper
experimental bounds by several orders of magnitude.
This problem can in principle be solved by constructing models
containing a hidden gauge group which  becomes strong at an
intermediate scale and confines the fractional charges into bound
states\cite{eln}. 

Finally, from the technical point of view, the greatest difficulty
in these constructions is to obtain a three generation unified or
partially unified model, which at the same time retains the  successful
low  energy predictions of the supersymmetric GUT's.  In fact, 
we know that using the higgs and fermion content of the  minimal
 supersymmetric standard model, the three gauge couplings $g_{1,2,3}$
 of the standard gauge group attain a common value at a scale 
$M_{GUT}\sim 10^{16}GeV$. However in strings, the unification point
($M_{string}$) is not an arbitrary parameter: it is a calculable quantity 
from the  first principles of the theory and at the one loop level
is found to be around two  orders of magnitude larger that $M_{GUT}$,
$M_{string}\sim 0.5 g_{string}\times 10^{18}GeV$. 
String threshold corrections \cite{thres} which can also be computed 
in terms of quantities related to the heavy string modes, do not bring
closer these two scales.
The consistency of string unification and low energy values of gauge 
couplings can be arranged if suitable extra matter representations
and  proper intermediate gauge group breaking steps are included.

A partially unified group which fulfills the basic requirements
\cite{alr}, is  based on the Pati-- Salam \cite{PS} gauge symmetry 
$SU(4)\times SU(2)_L \times SU(2)_R$. The symmetry can break down
to the standard model gauge group without using adjoint or any higher
representations. Color triplets and higgs doublets arise in different
representations, thus the model is free from doublet--triplet splitting
complications, as the triplets become massive from simple trilinear
couplings. There are no dangerous proton decay mediating  gauge
 bosons, thus the SU(4) breaking scale can be lower than the  GUT
scale predicted by other rival unified groups. Furthermore, a recent 
non -- renormalisable operator analysis\cite{steve} of its 
supersymmetric version,  has shown quite remarkable  features 
on the fermion mass matrices\cite{AK,elen}, which provide a strong
motivation to study the string derived model in more detail.
The renormalisation group analysis of the string version has already
been studied  in detail in many papers,  taking into account GUT, 
supersymmetric and string  threshold corrections\cite{alt,BL,homer}.  
It was shown  that it is possible to obtain the correct range of the 
low energy parameters while having two different  scales,  (a string 
$M_{string}\sim 10^{18}GeV$ and a ``GUT'' SU(4) gauge breaking  around
$(10^{15}-10^{16})GeV$) provided there is an intermediate scale $\sim
10^{10}GeV$ where some ``exotic'' states acquire their masses. This
was necessary to compensate  for the splitting of the three standard 
model coupling constants, caused by the different evolution of the 
$g_L,g_R, g_4$ gauge couplings  in the range $M_{string}- M_{GUT}$. 
However, a more natural way to achieve unification of the standard 
model gauge couplings at $\sim 10^{16}GeV$, is to include suitable 
representations which enforce the same (or even approximately
 similar) running of the $g_L,g_R, g_4$  couplings between 
$M_{string}-M_{GUT}$ \cite{steve}.

In the present work, we wish to present an alternative version of 
the  string model based on a different $b_{1,2,3}$ subset of basis
vectors.  This new construction offers some rather interesting 
features with respect to its predecessor: 
First, the fractionally charged states  appear now with non--trivial
transformation properties under a hidden gauge group (namely $Sp(4)$).
Although this is not probably enough to confine the fractional states
at a rather high scale, the above construction can be viewed as an
example  how to proceed for a more realistic model. 
Second, due to a symmetric appearance of the $L-R$-parts
of the various representations in this model, it is in
principle possible to obtain almost equal values of the  
$g_L,g_R,g_4$ couplings after their running down to $M_{GUT}$. 

Before we proceed to the derivation of the string model,  in
order to make clear the above remarks we  briefly start  with 
the basic features of the supersymmetric minimal version. Left 
and right handed fermions (including the right handed neutrino)
are accommodated in the  $(4,2,1)\; ,(\bar 4,1,2)$ representations 
respectively. Both pieces form up the complete $16^{th}$ 
representation of  $SO(10)$. The symmetry breaking down to the
standard model occurs in the presence of the two standard doublet
higgses  which are found in the $(1,2,2)$ representation of the 
original symmetry of the model. (The decomposition of the latter under 
the $SU(3)\times SU(2)\times U(1)$ gauge group results to the two
higgs doublets $(1,2,2)\ra (1,2,\frac 12 ) + (1,2,-\frac {1}{2})$.)
The $SU(4)\times SU(2)_R \ra SU(3)\times U(1)$ symmetry breaking 
is realized at a scale $\sim 10^{15-16}GeV$, with the introduction
of a higgs pair belonging to  $H + \bar{H}= (4,1,2) + (\bar 4,1,2)$
representations. 

The asymmetric form of the higgs fourplets
with respect to the two $SU(2)$ symmetries of the model, causes
a different running for the $g_{L,R}$  gauge couplings from the
string scale down to $M_{GUT}$. The possible existence of a 
new pair of representations with $SU(2)_L$ -- transformation 
properties (as suggested in \cite{steve})
which become massive close to $M_{GUT}$, could adjust
their running so as to have $g_L = g_R$ at $M_{GUT}$. Moreover,
a relatively large number ($n_D$) of sextet fields ($n_D \sim 7$)
remaining in the massless spectrum down to $M_{GUT}$, would also
result to an approximate equality of the above with $g_4$ coupling.
Obviously, the equality of the three gauge couplings $g_{4,L,R}$ 
at the $SU(4)$ breaking scale  $M_{GUT}$, is of great importance. 
In practice, this means that the three standard gauge couplings 
$g_{1,2,3}$ start running from  $M_{GUT}$ down to low energies, 
with the same initial condition. Thus, choosing $M_{GUT}\sim 10^{16}GeV$,
we are able to obtain the correct predictions for $sin^2\theta_W$
and $a_3(m_W)$. As a matter of fact, the intermediate gauge breaking
step gives us one more free parameter (namely $M_{GUT}$), thus having
obtained the desired string spectrum we are free to choose its value
in order to reconcile the high string scale $M_{string}$ with the
low energy data.
 

With the above observations in mind, we will attempt to obtain a
variant of the $SU(4)\times O(4)$ model which pretty much satisfies 
the above  requirements.
The subset of the  first five basis vectors we are using in our
construction, including the ($1, S$) sectors are the following 
\bea
\begin{array}{llllllll}
1&=&\{\psi^{\mu},&\chi^{1...6},&(y\bar y)^{1...6},&
(\omega\bar \omega)^{1...6}&;
&\bar {\Psi}^{1...5}\bar{\eta}^{123}\bar{\Phi}^{1...8}\}\\
S&=&\{\psi^{\mu},&\chi^{1...6},&0,...,0,&0,...,0 &;&0,...,0\}\\
b_1&=&\{\psi^{\mu},&\chi^{12},&(y\bar y)^{3456}, 
&0,...,0&;&\bar {\Psi}^{1...5}\bar{\eta}^{1} \}\\
b_2&=&\{\psi^{\mu},&\chi^{34},&(y\bar y)^{12},
&(\omega\bar \omega)^{56}&;&\bar {\Psi}^{1...5}\bar{\eta}^{2}\}\\
b_3&=&\{\psi^{\mu},&\chi^{56},&(y\bar y)^{1234}, 
&0,...,0&;&\bar {\Psi}^{1...5}\bar{\eta}^{3}\}
\end{array}
\eea

All world sheet fermions appearing in the vectors of the above basis
are assumed  to have {\em periodic} boundary conditions.  Those not 
appearing in each vector are taken with {\em antiperiodic} ones.
We follow the standard notation used in references\cite{aehn,alr,af}.
Thus, $\psi^{\mu},\chi^{1...6},(y/\omega)^{1...6}$ are real 
left, $,(\bar y/\bar \omega)^{1...6}$ are real right, and 
 $\bar {\Psi}^{1...5}\bar{\eta}^{123} \bar {\Phi}^{1...8}$ 
are complex right  world sheet fermions.
In the above, the  basis element $S$  plays the role of the
supersymmetry generator as it includes exactly eight left movers.
$b_{1,2}$ elements reduce  the  $N=4$ supersymmetries 
successively into $N=2,1$. 
Furthermore, the above set breaks the original symmetry of the  right
part down  to an $SO(10)$  gauge group  corresponding to the five 
($\bar {\Psi}^{1...5}$) complex world sheet fermions while all chiral 
families at this stage belong to the $16^{th}$ representation of the 
$SO(10)$. Note here the difference of the third basis element $b_3$ with 
the one used in previous constructions \cite{aehn,alr,af}.
To reduce further the $SO(10)$ symmetry to the desired 
$SO(6)\times O(4)$ gauge group, we introduce the  basis elements
$b_4 = \{(y\bar y)^{126}, (\omega\bar \omega)^{126};0,...,0\}$, 
$b_5=\{(y\bar y)^{136},(\omega\bar \omega)^{136};\; 0,...,0\}$
and the vector
\bea
\begin{array}{llllllll}
\alpha&=&\{0,&0,...,0,&(y\bar y)^{3},
&(\omega\bar \omega)^{3}&;&\bar {\Psi}^{123}\bar\eta^{123}
\bar {\Phi}^{1...6}\}
\end{array}
\eea
These three vectors complete our basis for the model under consideration.
In particular, the vector $\alpha$  breaks the original gauge  group to 
the following  symmetry 
\beq
\left[SO(6)\times SO(4)\right]_{obs.}
\times U(1)^3\times
\left[SO(12)\times Sp(4)\right]_{Hidden}
\label{eq:sym}
\eeq
$SO(6)\sim SU(4)$ corresponds to the three complex fermions 
$\bar {\Psi}^{123}$, while $\bar {\Psi}^{45}$  generate the $O(4)\sim
SU(2)_L\times U(2)_R$ part of the observable gauge symmetry.
$SO(12)$ corresponds to $\bar{\Phi}^{1...6}$ while $SO(5)\sim Sp(4)$
to $\bar \omega \bar{\Phi}^{78}$.
We have introduced subscripts to denote the  {\em observable} and 
{\em Hidden}  part of the symmetry. 
A well known feature of these constructions is the appearance of
various $U(1)$ factors (three in the present case ) which act as a 
family symmetry \cite{IR} between the generations.  As we will
see soon, the fractionally charged states in the observable sector
belong also to the $4 = \bar 4$ representations of the $ Sp(4)
\sim SO(5)$. The particular content of the model depends also on
the choice of the  specific set of the projection coefficients
$c\left[{}^{b_i}_{b_j}\right]=e^{i\pi c_{ij}}$. In order to guarantee
the existence of $N=1$ space time supersymmetry, we choose
$c\left[{}^{S}_{b_j}\right]= 1$ for $j=1,2,3$, while for the other
coefficients one possible choice  is 
$c\left[{}^{a}_{a}\right] = c\left[{}^{b_i}_{b_i}\right]=1$ for $i=4,5$,
$c\left[{}^{b_j}_{b_j}\right]= - 1$ for $j=1,2,3$
and $c\left[{}^{b_i}_{b_j}\right]= - 1, j>i$, while all the others
are fixed by the modular invariance constraints.

We start first by presenting the spectrum with the representations 
which are going to be interpreted as fermion generations and  $SU(4)$
breaking higgses.  Fermion generations arise  from $b_{1,2,3}$ sectors 
 appearing in symmetric representations under the $SO(6)\times O(4)$
symmetry. Thus it makes no difference which of the two resulting 
representations of  $b_{1,2,3}$  will  accommodate the left or right 
components  of the fermion generations. The choice of the assignment
however, is crucial for  the higgs fourplets which are not symmetric 
under  the two $SU(2)$'s.  Thus, starting with one of the two possible 
choices the sectors which provide with the fermion generations and 
possible $SU(4)$ breaking higgses are{\footnote{
The second case arises by interchanging $4\lra \bar 4$ ,  
$2_L\lra 2_R$ in the above sectors, and will be commented below.}}

\begin{eqnarray} 
\begin{array}{llllllll}
b_1 : & F_1 & = & (  4,2,1)_{(-1/2,0,0)}&
;&{\bar F}_1 & = & ( \bar 4,1,2)_{(1/2,0,0)}
\\
b_2 : &F_2 & = & ( 4,2,1)_{(0,-1/2,0)}&
;&{\bar F}_2 & = & ( \bar 4,1,2)_{(0,1/2,0)}
\\
b_3 : & F_3 & = &(4,2,1)_{(0,0,-1/2)}&
;&\bar F_3& = & (\bar 4,1,2)_{(0,0,-1/2)}
\\
b_2+b_4 : & F_{24} & = & ( 4,2,1)_{(0,1/2,0)}&
;&\bar F_{24} & = & (\bar{4}, 1, 2)_{(0,-1/2,0)}
\\
 b_3+b_4+b_5 : &\bar{F}_{345} & = & ( \bar{4},2,1)_{(0,0,1/2)}&
 ;&\bar{F}'_{345} & = & (\bar{4}, 1, 2)_{(0,0,-1/2)}
\end{array}
\end{eqnarray} 

The above  representations  of the observable sector transform trivially
under the hidden gauge group. However, they all appear  charged under
the three $U(1)$ factors corresponding to $\bar\eta_1,\bar\eta_2,
\bar\eta_3$ world--sheet  fermions. These charges are denoted with 
the three  indices  in the above representations. 
$F_{1,2,3}, \bar{F}_{1,2,3} $ can accommodate the three generations,
while from the $(b_2+b_4)$ and $ (b_3+b_4+b_5)$  sectors we get
a pair of family -  antifamily ($F_{24}-\bar{F}_{345}$) left--fourplets.
Unfortunately, in  this case the two remaining representations 
$\bar{F}'_{345},\bar F_{24}$ cannot play the role of  the two $SU(4)$
higgses, as they are both of the type  $\bar{H}_{1,2}=(\bar{4},1,2)$. 
More over, this spectrum apparently creates an anomaly with respect to 
the $SU(4)$  gauge group, since there is an excess of fourplet over anti
-- fourplet fields; however,  there  is a pair of exotic states 
$( 4,1,1)^n_{(1/2,0,0)} +  ( 4,1,1)^{\bar n}_{(1/2,0,0)}$ 
 with fractional charges arising from  the  sector
 $(1+b_2+b_3+b_4+\alpha )$ which guarantee the anomaly cancellation.
The novel feature of these representations here, is their non -- trivial 
transformation under part of the hidden non -- abelian gauge group. 
In fact they  belong to the $n = \bar n = 4$ representation (denoted as 
superscript) of the $Sp(4)$ symmetry. As we will see soon,
 this is also true for the rest of the exotics in this construction. 
Provided the hidden group confines at some later stage, this allows for 
the  possibility of forming various types of condensates. By choosing
proper flat directions, such states may become massive and disappear 
from the light spectrum, while some of them can have the right higgs 
properties so that they can be used to break the $SU(4)$ symmetry. 
Indeed, in order to examine this case further,  in the following 
let us continue with the relevant representations.
{}From the sectors $(1+b_1+b_2+\alpha )$, ($1+b_1+b_2+b_4+b_5+\alpha$)
and $(1+b_2+b_3+b_4+\alpha )$ we obtain six pairs of ``exotic'' doublet 
states  $(1,1,2)^{(n/\bar n)}+(1,2,1)^{(\bar n/ n)}$, possessing half 
integer ($\pm 1/2$) electric charges. Interestingly enough, these exotic
states can in principle condense with the $( 4,1,1)^n_{(1/2,0,0)} +  
( 4,1,1)^{\bar n}_{(1/2,0,0)}$ states into the missing higgs fourplets
$H_{1,2} = ( 4,1,2)$ at a later scale. (Their $U(1)$ -- charges
depend on the specific $(1,1,2)$ representations). Thus in this way 
there can exist now two higgs pairs (namely $H_{1,2}+\bar{H}_{1,2}$)
where either of them can break the $SU(4)$--symmetry to the standard
model.  However, of crusial importance is the confinment scale $M_C$ 
of the $Sp(4)$ symmetry, as it simultaneously defines the $SU(4)$ 
breaking scale of the observable symmetry. 
This can be calculated from the formula
\begin{equation}
M_C = M_{string}
 Exp\{\frac{2\pi}{b_{SO_5}}
(\frac{1}{\alpha_{string}}-\frac{1}{\alpha_{c}})\}
\end{equation}
where $b_{SO_5} = -3 C_2(SO_5) + 2 n_4 + n_2$ is the beta function of
 $SO(5)$, while $C_2(SO_5) = 3$. For two fourplet higgses we need
$ n_4 = n_2 = 2$ thus $b_{SO_5} = - 3$ as in the case of the $SU(3)$,
which means that the confining scale is rather low. However, there
are some important differences which should be  mentioned.
First, the initial scale where the renormalisation starts is 
$M_{string}$ which is two orders higher than the supersymmetric
unification scale $M_{GUT}$. Furthermore the unified coupling 
$a_{string}$ turns out to be larger than the common gauge coupling
$a_{GUT}$ in the minimal supersymmetric unification. For example in
\cite{..} it is found $a_{string}\sim 1/20$, while $a_{GUT}\sim 1/25$.
Thus, in contrast to the $SU(3)$, for the Sp(4) confining scale
one finds $M_{Sp_4}\sim 10^7 GeV$. This scale is still rather low
compared to the usual grand unification. However, in the case of
the $SU(4)$ `unification' this is not a disaster; as we have already
pointed out, there are no gauge bosons mediating proton decay, thus
a low energy breaking scale is not necessarily in contradiction with
the  low  energy phenomenology. Nevertheless, it would be desirable
to obtain a rather higher confinement scale close to the `conventional'
minimal supersymmety unification point $\sim 10^{15-16}GeV$. This of
course would need a confining group with  rank higher that the $Sp(4)$.

{}From the Neveu-Schwarz sector we get the following fields: Two higgs
 fields of the type $(1,2,2)_{(0,0,0)}$ under the observable  $SU(4)
\times SU(2)_L\times SU(2)_R$ gauge group, and no charges under the three 
family--type $U(1)$ symmetries. Six sextet fields
$(6,1,1)_{ (\pm 1,0,0) + perm.}$ 
Various singlet fields $\chi^{\alpha}_{(\pm 1, 0, \pm 1)},
\chi^{\beta}_{(\pm 1, \pm 1 ,0)}, \chi^{\gamma}_{(0,\pm 1, \pm 1) }$ 
with integer $(\pm 1)$ surplus $U(1)$ charges are also available.
Representations with the same transformation properties but different
charges under the three $U(1)$-- family symmetries are obtained from the 
sectors $S+b_2+b_3$, $S+b_1+b_3$ and $S+b_1+b_3+b_4$. In particular, 
they give singlet fields analogous to those of the NS -- sector but 
with  half -- integer  extra $U(1)$ charges, 
$\xi^{\alpha}_{(\pm 1/2, 0, \pm 1/2)},
\xi^{\beta}_{(\pm 1/2,\pm 1/2,0)}, \xi^{\gamma}_{(0,\pm 1/2,\pm 1/2 )}$,
 and $\Sigma_{(\pm 1,\pm 1/2,\pm 1/2)}$.  In addition
 in the massless spectrum there exist vector representations of the 
hidden part of the symmetry which do not have transformation properties
under the observable gauge group. Thus, each of the above three sectors 
gives the ${\bf 12}$ of $SO(12)$ and ${\bf 5}$ of $SO(5)$. The resulting 
three  $12^{th}$ irreps do not play any role in the observable world,
however if the ${\bf 5} 's$ remain massless, they can lower dangerously
 the  confining  scale. 
Finally, from the same sectors one gets sextet fields $D_{1,2,3} =
 (6,1,1)_{(0,1/2,1/2)}$, $(6,1,1)_{(\pm 1/2,0,1/2)}$ and higgses 
$h_{1,2,3} = (1,2,2)_{(0,-1/2,1/2)}$, $(1,2,2)_{(\pm 1/2,0,1/2)}$. 
At least one of the latter is expected to acquire a vacuum expectation
value (vev) along its two neutral components in order to give masses to 
fermion generations through
Yukawa couplings allowed by gauge and string symmetries. Although only
few couplings are expected to be present at the trilinear superpotential,
there is a large variety of singlet fields possessing various $U(1)$
charges which are going to form non -- renormalisable mass terms.

Let us briefly now discuss the fermion masses. Light fermions acquire 
their masses with the usual higgs mechanism, when some of the $(1,2,2)
\ra (1,2,\frac 12 ) + (1,2,-\frac {1}{2} )$ higgs representations 
develop vevs.  If we assume that below $M_{GUT}$ the model behaves
approximately as the minimal supersymmetric  standard model,  only 
one pair of the available electroweak higgs doublets
(or only a linear combination of them ) should remain light.
Then, in the trilinear superpotential,
a coupling of the form $\lambda^0_{ijk}F_{i}\bar F_{j} h_{k}$ will 
provide with masses  the fermions of the third generation, 
with the GUT--predictions $m_t^0 = m_{\nu_D}^0$, $m_b^0 = m_{\tau}^0$,
where  $m_{\nu_D}^0$ is the Dirac neutrino mass. A remarkable  feature 
of these string  models is the  generic prediction that the Yukawa coupling
$\lambda_t^0$ responsible for the top-quark mass is large and of the same 
order  with  the common  gauge coupling  at the string scale,
$\lambda_t^0 = \sqrt{2}g_{string}$, leading to a top mass of the
${\cal O}(180)GeV$\cite{..}. This is compatible with previously 
proposed SUSY-GUT models which predicted  radiative symmetry breaking
and  a large top mass  with a single third  generation Yukawa 
coupling\cite{acw}.
The  bad $(m_t^0, m_{\nu_D}^0)$ relation is handled with the
``see-saw''-- type relation through a term of the form  $ H \bar F_i\Phi_n 
\ra  <H>\nu_{R_i} \Phi_n $ as described in previous works\cite{alr,elen}.
The rest of the entries of the fermion mass matrices are expected 
to fill up when non--renormalisable contributions to the
superpotential are taken into account. Additional colored triplets
$d^c_H,\bar{d^c}_H$ remaining from the $H+\bar H$ representations form 
massive states with $D_3,\bar D_3$ states arising from the decomposition
of the sextet fields $D \ra D_3+\bar D_3$, through terms of the form
$H H D , \bar H\bar H D$\cite{alr}. Note that some of them could be
harmless even if they get mass at a relatively low scale $\sim 10^7 GeV$ 
provided they do not couple with the ordinary matter at the tree level.

Finally, the family ${F}_{24}=( 4, 2,1)$ -- antifamily $\bar{F}_{345}= 
(\bar 4, 2,1)$  pair 
can become massive either at the tree level or from a higher order
non--renormalisable coupling of the form ${\cal W} \; \supset \; \;
<\Phi_i>( 4, 2,1)(\bar 4, 2,1)$, with $<\Phi_i>\sim M_{GUT}$. In fact 
the singlet vevs are not completely arbitrary in these constructions.
From the three family type $U(1)$'s of the present model, one can define
two linear combinations (say $U(1)_1-U(1)_2-U(1)_3, U(1)_2-U(1)_3$) which
are anomaly free, while the remaining orthogonal combination remains 
anomalous. The latter  is broken by the Dine-Seiberg-Witten mechanism
\cite{dsw} in which a potentially large supersymmetry--breaking D--term 
is generated, by the vacuum expectation value of the dilaton field.
To avoid this situation, one has to choose a D-- and F-- flat direction 
in the scalar potential by assigning proper vevs to some of the scalar 
fields. The natural scale of these singlet vevs turns out to be
 $M_{string}\ge <\Phi_i>\ge M_{GUT}$.

Let us finally analyse  the alternative accommodation of the  fermion 
generations and higges under the observable symmetry. This can be
easily obtained by  interchanging $4\lra \bar 4$ and $2_L\lra 2_R$ 
in the relevant sectors. The three sectors $b_{1,2,3}$ provide
 again the  three generations. From  $b_2+b_4$ one gets  $F_{24}  =  
( 4,2,1) , \bar F_{24}   =   (\bar 4, 1, 2)$ while $ b_3+b_4+b_5 $ 
gives   $F_{345}  =  ( 4,1,2)  , F'_{345}  = ( 4, 2, 1) $. 
(of course the $U(1)$ charges are not affected). Thus, now 
the  higgs fourplets  $H +\bar H$  needed to break the $SU(4)$ symmetry 
are contained in the $(b_2+b_4)$ and $ (b_3+b_4+b_5)$  sectors. In fact 
we  can now identify  $ H \equiv F_{345}=(4,1,2)$ and  $\bar H  \equiv 
\bar{F}_{24} =(\bar{4},1,2)$.  It is possible however that a detailed 
phenomenological analysis  of the model would require some linear
combinations  of $F_i$'s and $\bar F_i$'s to be  interpreted as the 
$SU(4)$ breaking higgses of the model. Thus, in this case the  higgs 
particles are not formed by condensates, therefore the `GUT' scale
is not related to the confinment scale. We may choose then $M_{GUT}\sim
10^{15-16}GeV$ and obtain a renormalisation group running of the gauge
couplings as described above. The present accommodation however, creates
a new problem; the two remaining pieces of $(b_2+b_4)$ and $ (b_3+b_4+b_5)$ 
sectors, have the same transformation properties with the left handed 
fermion generations. These two remaining $(4,2,1)$ states are rather 
difficult to become massive. However, it is possible that after the
$SU(4)$ breaking the resulting  colored triplets and doublets may combine 
with their conjugate partners arising from the composite states 
$(\bar 4, 2,1)$ (which now tranform as anti -- fourplets under the 
interchange $4\lra\bar{4}$) through non--renormalisable
terms resulting in an effective mass term much lower than the scale 
$M_{GUT}$. 


The above model, is not of course a fully realistic model for the low energy 
theory. However, it is a rather interesting improvement of a previous version 
which was based on the same gauge symmetry. Its advantages with respect to 
the old version can be briefly summarized in the  following  points: 
 Fractionally charged states transform non trivially under a hidden 
gauge group (namely $Sp(4)$) which forces them to form bound states.
Specific composite states can play the role of the higgses  which 
break the $SU(4)\times SU(2)_R$ symmetry while the most of the remaining  
hopefully may combine in various terms with other fields into relatively
heavy  massive states escaping detection by the present experiments.
The main drawback of this construction is that the $Sp(4)$ group 
falls rather short to confine these charges at a suitably high scale. 
  A novel feature of this construction of the model is also the 
choice of the  vectors $b_{1,2,3}$ which are different than the already 
used in the flipped $SU(5)$\cite{aehn} and  standard model\cite{af}
constructions. Since the previous  $SU(4)$  model has been pretty  much 
similar to the flipped $SU(5)$ we think  that the three basis  vectors 
$b_{1,2,3}$ used here, can also offer new  possibilities for these  
constructions which are worth exploring.

\end{document}